\documentclass[%
twocolumn,
superscriptaddress,
 amsmath,amssymb,
 aps,
pra,
]{revtex4-2}

\usepackage{graphicx}% Include figure files
\usepackage{dcolumn}% Align table columns on decimal point
\usepackage{bm}% bold math
\usepackage{color}
\usepackage{xcolor}
\usepackage{ulem}
\usepackage{stackengine}
\usepackage{braket}
\usepackage{multirow}
\usepackage[whole]{bxcjkjatype}% Using Japanese with pdflatex
\newcommand{\figref}[1]{Fig.~\ref{fig:#1}}

\newcommand{\equref}[1]{Eq.~(\ref{eq:#1})}

\begin{document}
%\preprint{APS/123-QED}

\title{
Localized states of the non-Hermitian Schr\"odinger equation
with a complex $1/r^{2}$ potential: from Efimov states to bipolarons
}% 
\author{Kyoka Tsunoda}
\email{kyoka.tsunoda@uec.ac.jp}
\affiliation{Department of Engineering Science, University of Electro-Communications, Chofu, Tokyo 182-8585, Japan}
\author{Kazuki Yamamoto}
\email{kazuki-yamamoto@omu.ac.jp}
\affiliation{Research Institute for Innovation and Co-Creation, Osaka Metropolitan University, Sakai, Osaka 599-8531, Japan}
\affiliation{Department of Physics, Osaka Metropolitan University, Sumiyoshi, Osaka 558-8585, Japan}
\affiliation{Nambu Yoichiro Institute of Theoretical and Experimental Physics (NITEP), Osaka Metropolitan University, Sumiyoshi, Osaka 558-8585, Japan}
\author{Shimpei Endo}
\email{shimpei.endo@uec.ac.jp}
\affiliation{Department of Engineering Science, University of Electro-Communications, Chofu, Tokyo 182-8585, Japan}
\affiliation{Institute for Advanced Science, University of Electro-Communications, Chofu, Tokyo 182-8585, Japan}

\date{\today}

\begin{abstract}
We study the non-Hermitian Schr\"odinger equation with a complex $1/r^{2}$
potential, which bridges Hermitian few-body and non-Hermitian many-body
physics: a real attractive potential describes Efimov bound states, whereas a purely imaginary one describes the medium-induced interaction between two heavy polarons at large separations. 
We solve this equation analytically, obtaining both the localized and continuum states, and compare the solutions with numerical and WKB results.  
As the imaginary part of the potential is increased, the energies of the Efimov bound states acquire imaginary parts and rotate counterclockwise in the complex plane. 
On approaching the purely imaginary potential, the excited states disappear sequentially, while the Efimov ground state is smoothly connected to a bipolaron state localized at short distance.
Such a bipolaron is localized solely by the medium-induced imaginary interaction, without any attractive real potential. 
Solving the equation by finite differences, we find that each bipolaron state is necessarily accompanied by another localized state that oscillates on the scale of the mesh size and hence has no counterpart in the continuum limit. 
Each such pair coalesces at an exceptional point as the strength of the purely imaginary potential is varied, whereas the coalescence is lifted once a finite real part is introduced. We show that this behavior originates from a symmetry of the discretized Schr\"odinger equation that holds for any purely imaginary potential.
\end{abstract}

\maketitle

%\tableofcontents
%************************************************************************
\section{Introduction}

\vspace{0.3cm}
Extending quantum problems into the complex domain often reveals phenomena absent on the real axis. 
Prime examples are resonant states, whose complex energies describe the decay of unstable nuclei and atoms~\cite{Gamow1928,Siegert1939,Newton1982}. More recently, the Hamiltonian itself has been extended to be non-Hermitian~\cite{heiss12,okuma20,kawabata19,ashida20,bergholtz21}. 
Typically, a non-Hermitian Hamiltonian provides an effective description of open quantum systems upon postselecting special measurement outcomes~\cite{daley14}. 
The interplay of dissipation and correlations gives rise to rich phenomena absent in Hermitian systems, such as fermionic superfluidity accompanied by exceptional points~\cite{yamamoto19,takemori24,takemori25}, $\mathcal{PT}$-symmetry-enriched localization-delocalization transitions~\cite{hatano96,hatano97,hatano98,fukui98,hamazaki19}, anomalous renormalization group flows in Kondo effects~\cite{nakagawa18,lourenco18,burke25,yamamoto25}, and critical phenomena in dissipative Tomonaga-Luttinger liquids~\cite{ashida16,ashida17,yamamoto22,yamamoto23}.

Complex parameters also appear in the Efimov effect~\cite{efimov1970energy,kraemer2006evidence,PhysRevLett.112.190401,science.1182840,PhysRevLett.112.250404,PhysRevLett.113.240402,Braaten2006259,naidon2017efimov,RevModPhys.89.035006,d2018few,endoepelbaumQcl2024}, where three particles with resonant two-body interactions form a geometric tower of weakly bound three-body states.
The exchange of the third particle between the other two renders the hyperangular eigenvalue imaginary, which manifests itself as an attractive $1/R^{2}$ potential in the hyperradius $R$~\cite{efimov1970energy,Braaten2006259,naidon2017efimov}.
A quantum anomaly breaks the continuous scale invariance of this attraction down to a discrete one.
The Efimov states have been observed in cold atoms~\cite{kraemer2006evidence,PhysRevLett.112.190401,science.1182840,PhysRevLett.112.250404,PhysRevLett.113.240402}, where their discrete scale invariance
has been confirmed, and in $^{4}$He clusters~\cite{kunitski2015observation}. They have also been studied in nuclear~\cite{hoyle1954nuclear,AnnRev_HamPlatt}, hadronic~\cite{PhysRevLett.91.102002}, and magnetic systems~\cite{nishida2013efimov}.

Mediated interactions also play an important role in quantum many-body systems.
A prominent example is the phonon-mediated attraction between electrons, which underlies the mechanism of superconductivity~\cite{PhysRev.108.1175,PhysRev.167.331}.
Similarly, a medium induces an interaction between impurities immersed in it. Each impurity is dressed by the medium and forms a quasiparticle called a polaron~\cite{Landau1933,Pekar1946,
Frohlich1954,Feynman1955}, which appears across various fields, such as the Kondo effect~\cite{10.1143/PTP.32.37}, the quark-gluon plasma~\cite{PhysRevD.10.2445,PhysRevD.24.450}, neutron stars~\cite{PhysRevC.47.1077,tajima2024intersections},
semiconductors~\cite{sidler2017fermi,PhysRevX.13.031036}, and cold atoms~\cite{PhysRevLett.102.230402,Kohstall_2012,Koschorreck_2012, PhysRevLett.117.055301,PhysRevLett.117.055302,PhysRevLett.118.083602,PhysRevLett.122.093401,Yan_2020,Cetina_2016,PhysRevResearch.7.L022002,Massignan_2014}.
The interaction between two polarons has been actively studied as a relevant step toward understanding many-body phases of quantum matter, with well-known examples including the RKKY interaction~\cite{PhysRev.96.99,10.1143/PTP.16.45,PhysRev.106.893} in heavy-fermion materials~\cite{DONIACH1977231} and the Yukawa~\cite{Pethick_Smith_2008} and van der Waals~\cite{PhysRevLett.129.233401} forces between impurities in a Bose-Einstein condensate.
Such induced interactions between polarons have recently been measured in cold atoms~\cite{DeSalvo:2019juk,Baroni:2023ejp,PhysRevLett.131.083003,w7rd-2qpv} and semiconductors~\cite{muir2022interactions,PhysRevX.13.031036}.

A polaron is a prototypical open quantum system immersed in a surrounding medium, and its dissipative dynamics has also been studied from this
perspective~\cite{Lampo_2017,PhysRevA.98.063630}.
Recently, the induced interaction between two heavy polarons has been shown to acquire an imaginary $1/r^{2}$ tail at low but finite temperature~\cite{PhysRevA.110.033304}. Since this imaginary potential is much longer ranged than the real part~\cite{PhysRev.96.99,10.1143/PTP.16.45,PhysRev.106.893}, it dominates the long-range, low-energy dynamics of the two polarons.
Although this complex potential has been derived microscopically, its consequences for the two-polaron problem remain largely unexplored.

In this paper, we solve the non-Hermitian Schr\"odinger equation with a
complex $1/r^{2}$ potential to reveal the behavior of two heavy polarons in a dissipative medium, and clarify how their
localized states are connected to the Efimov states by continuing the potential to a real attractive one.
We obtain analytical expressions for the energies and the wave functions of the localized states as well as of the continuum states, and corroborate them with the WKB approximation and with numerical solutions obtained by the finite-difference method. 
As the imaginary part of the potential is increased, the energies of the Efimov states acquire imaginary parts and rotate counterclockwise in the complex energy plane, with the states continued from the higher excited Efimov states disappearing at weaker imaginary potentials. A localized state surviving up to the purely imaginary potential is a bipolaron state, which is formed solely by the dissipative interaction induced by the medium. The Efimov state and the bipolaron state are thus continuously connected in the complex plane, providing a link between few-body physics in vacuum and two-body physics in a dissipative medium. 
We further find that, in the finite-difference calculation, each bipolaron state is accompanied by another localized state that oscillates on the scale of the mesh spacing. The two states can form an exceptional point when the potential is purely imaginary, as a consequence of a symmetry of the discretized Hamiltonian.

%************************************************************************

\section{\label{sec:model}Model and Results}
\subsection{Localized and continuum states in the complex $1/r^{2}$ potential}

We study the non-Hermitian Schr\"{o}dinger equation with a complex parameter $s^{2} \in \mathbb{C}$,
\begin{equation}
    \label{eq:hamiltonian_1}
    \left[-\frac{1}{2\mu}\frac{d^{2}}{dr^{2}}
    + \frac{s^{2}-1/4}{2\mu r^{2}}\right]\psi(r) = E \psi(r),
\end{equation}
where $\mu$ is the reduced mass of the two particles. Throughout this paper, we use natural units with $\hbar=1$. For real positive $s^{2}$, in particular $s^{2}=(\ell+1/2)^{2}$, Eq.~(\ref{eq:hamiltonian_1}) reduces to the free radial equation with angular momentum $\ell$ and admits no bound states. For real negative $s^{2}$, in contrast, \equref{hamiltonian_1} can support a series of bound states, known as the Efimov states~\cite{kraemer2006evidence,efimov1970energy,Braaten2006259,naidon2017efimov,RevModPhys.89.035006,d2018few,endoepelbaumQcl2024}. Indeed, close to the unitary limit, the hyperradial equation obtained after solving the three-body hyperangular equation takes the form of \equref{hamiltonian_1} with a purely imaginary hyperangular eigenvalue $s = i|s|$. The resulting spectrum obeys the discrete scale invariance, $E_{n+1}/E_{n} = e^{-2\pi/|s|}$, with the overall scale fixed by a short-distance boundary condition $\psi(r=r_\mathrm{min}) = 0$. The Efimov states and their discrete scale invariance have clearly been observed in cold-atom experiments~\cite{kraemer2006evidence,PhysRevLett.112.190401,science.1182840,PhysRevLett.112.250404,PhysRevLett.113.240402}.

While the Hamiltonian in Eq.~(\ref{eq:hamiltonian_1}) is Hermitian for real $s^{2}$, it becomes non-Hermitian once $s^{2}$ acquires an imaginary part.
Inverse-square interactions have also been studied in other non-Hermitian contexts, including a Calogero--Sutherland model with a complex interaction describing dissipative Tomonaga--Luttinger liquids~\cite{yamamoto23}, a non-Hermitian Klein--Gordon field with a real inverse-square coupling and an absorbing boundary condition at the singular point~\cite{Haghighat2026}, and non-Hermitian random matrices with Calogero scattering states as eigenvalue densities~\cite{Xiao2026}.
In the present work, the complex parameter $s^{2}$ in Eq.~(\ref{eq:hamiltonian_1}) characterizes the two-body interaction between polarons immersed in a quantum degenerate bosonic or fermionic medium. Indeed, Ref.~\cite{PhysRevA.110.033304} demonstrates that the imaginary part of the potential between the two heavy impurities behaves as $-1/r^{2}$ at large distances in finite-temperature bosonic and fermionic media. Notably, the imaginary part is much longer ranged than the real part of the potential, which falls off as $1/r^{3}$ in fermionic~\cite{PhysRev.96.99,10.1143/PTP.16.45,PhysRev.106.893} and as $1/r^{6}$ to $1/r^{7}$ in bosonic media~\cite{PhysRevLett.129.233401}. This effective description~\cite{PhysRevA.110.033304}, valid at long times and for weak impurity-medium coupling, justifies studying the low-energy (i.e., large-distance) behavior of the two heavy polarons via the dissipative potential in \equref{hamiltonian_1} with $\mathrm{Im}[s^{2}]<0$. The energy then acquires a negative imaginary part representing the decay rate of the two-polaron state due to collisions with the bath particles. Without loss of generality, we take the branch $\pi/2 < \arg s < \pi$.

We solve \equref{hamiltonian_1} under the boundary conditions $ \psi(r=r_\mathrm{min}) = \psi(r=r_\mathrm{max}) = 0$. The short-distance boundary condition at $r_\mathrm{min}$ is essential for avoiding the Thomas collapse~\cite{PhysRev.47.903,efimov1970energy,Braaten2006259,naidon2017efimov,RevModPhys.89.035006,d2018few}, which is caused by the rapid oscillation of the wave function as $r\rightarrow 0$. More physically, 
$r_\mathrm{min}$ should be understood as a matching scale beyond which the asymptotic $1/r^2$ form of the potential is applicable.
The physics at shorter distances is encoded in $r_\mathrm{min}$, which serves as a quantum defect parameter~\cite{PhysRevA.19.1485,PhysRevA.78.012702,PhysRevA.64.022717,PhysRevA.84.042703,PhysRevA.110.033305,PhysRevResearch.7.023162,Jie_2025,8x96-ygjt} characterizing the short-range phase shift arising from the direct $s$-wave scattering between the heavy impurities and from the real part of the medium-induced potential between the polarons~\cite{PhysRev.96.99,10.1143/PTP.16.45,PhysRev.106.893,PhysRevLett.129.233401}. The large distance boundary condition at $r_\mathrm{max}$ is required to describe the continuum states, whereas it is irrelevant for localized states concentrated around short distances $r \sim r_\mathrm{min}$, as we show below.

%************************************************************************
\begin{figure*}[!t]
    \centering
    \includegraphics[width=1.0\linewidth]{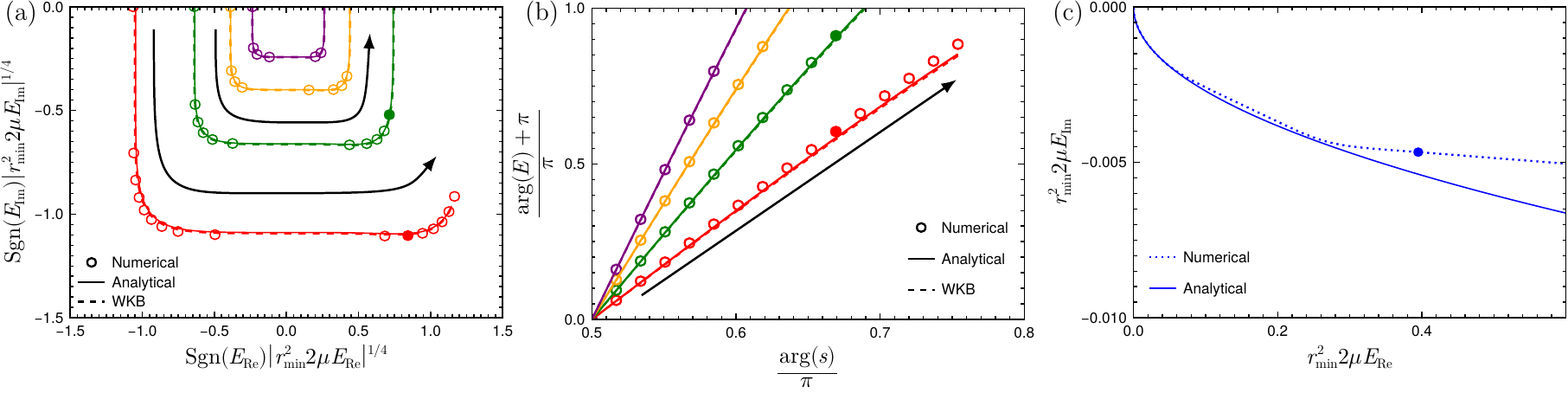} 
\caption{(a) Energies of the ground ($n=1$, red), first excited ($n=2$, green), second excited ($n=3$, yellow), and third excited ($n=4$, purple) localized states for the complex potential with $|s| \approx 3.163$ fixed. $\arg s$ is varied from $\pi/2$, corresponding to the Efimov state, to $\pi - \tfrac{1}{2}\arccos[1/(4|s|^{2})] \approx 0.754\pi$, where the potential becomes purely imaginary, $s^2 -1/4 = -10.0 i$. Arrows indicate the direction of increasing $\arg s$. The curves show the analytical results in \equref{energy_bound_state_analytical} (solid), WKB results in \equref{energy_bound_state_WKB_1} (dashed), and the circles show the numerical results with $r_\mathrm{min}=0.1$, $r_\mathrm{max}=100$, and $N=5000$, with $\arg s$ increased in 15 equal steps. (b) Argument of the complex energies as a function of $\arg s$. (c) Complex energies of the continuum states obtained analytically in \equref{energy_continuum_state_analytical_2} (solid) and numerically (dotted). The filled circles are $\arg[s] \approx 0.669 \pi$, for which the eigenfunctions are plotted in \figref{complex_pot_localized_state_wavefunc_1}.}
\label{fig:complex_pot_localized_state_energy_1}
\end{figure*}

Since \equref{hamiltonian_1} for real negative $s^{2}$ takes the same form as the hyperradial equation for Efimov states~\cite{kraemer2006evidence,efimov1970energy,Braaten2006259,naidon2017efimov,RevModPhys.89.035006,d2018few,endoepelbaumQcl2024}, its localized solutions can be obtained analytically in the same manner for complex $s^{2}$ as well. Indeed, by transforming \equref{hamiltonian_1} into the Bessel differential equation, its general solution is obtained as ($E = k^2 / 2\mu$)
\begin{equation}
    \label{eq:wavefunc_hankel_1}
    \psi(r) = \sqrt{r}\left[c_1 H_s^{(1)}(kr) + c_2 H_s^{(2)}(kr) \right],
\end{equation}
where the two independent solutions $H_s^{(1)}$ and $H_s^{(2)}$ are the Hankel functions with complex index $s$, defined by the Bessel functions $J_s$ and $Y_s$ as $H_s^{(1,2)}(kr) = J_s(kr) \pm iY_s(kr)$. Requiring that the wave function vanishes at large distance $\psi(r=r_\mathrm{max})\simeq \psi(r=\infty) =0$, we find $\psi(r) = c_1\sqrt{r} H_s^{(1)}(kr)$. This amounts to discarding the exponentially growing solution $H_s^{(2)}$, and keeping only the exponentially decaying solution $H_s^{(1)}$. Using the asymptotic form of the Hankel function $H_s^{(1)}$ in the low-energy limit $|kr_\mathrm{min}|\to 0$, valid for $-1<\mathrm{Re}[s]<1$, the short-distance boundary condition $\psi(r=r_\mathrm{min}) = 0$ leads to the transcendental equation $\left(kr_\mathrm{min}/2i\right)^{2s} = -\Gamma(s)/\Gamma(-s)$ (see Appendix~\ref{app:Analytic_derivation} for the derivation). This yields the analytical expression of the complex energy as
\begin{equation}
    \label{eq:energy_bound_state_analytical}
    E = \frac{1}{2\mu}\left(\frac{2}{r_\mathrm{min}}\right)^2 \exp\left[\pi i + \frac{1}{s} \ln \left(\frac{-\Gamma\left(s\right)}{\Gamma\left(-s\right)}\right)\right]\exp\left[ - \frac{2\pi i}{s}n\right],
\end{equation}
where $n$ is an arbitrary integer satisfying the low-energy condition $|E| \ll 1/2\mu r_\mathrm{min}^2$, as well as the condition $\mathrm{Im}[E]\le 0$ discussed later. This analytical expression for the complex energies provides a natural extension of the Efimov states to complex $s^{2}$. Indeed, for a real attractive potential with $s^{2}<0$, i.e., purely imaginary $s= i |s|$, \equref{energy_bound_state_analytical} yields negative real energies and reproduces the geometric series of Efimov bound states obeying the discrete scale invariance, $E_{n+1}/E_{n} = e^{-2\pi/|s|}$. For a non-Hermitian Hamiltonian with complex $s^{2}$, \equref{energy_bound_state_analytical} gives the complex energies of the localized states labeled by the principal quantum number $n$. They are analytically continued from the Efimov bound states at $s = i|s|$ as an imaginary part is added to the potential.

In \figref{complex_pot_localized_state_energy_1} (a), the solid curves show \equref{energy_bound_state_analytical} for $n=1$ to $n=4$ as $s$ is varied continuously in the complex plane from the Efimov limit $\arg s = \pi/2$ to the purely imaginary potential $s^2 - 1/4 = -i\beta$, with $|s|$ fixed. 
In the Efimov limit, where $s$ is purely imaginary and the Hamiltonian is Hermitian, \equref{energy_bound_state_analytical} gives the discrete-scale-invariant Efimov states on the negative real energy axis, with $n=1$ the ground state and $n\geq 2$ the excited states.
As $\arg s$ is increased, the states continued from these Efimov states acquire finite lifetimes $\mathrm{Im}[E]<0$ and rotate counterclockwise in the complex energy plane, with $\mathrm{Re}[E]$ turning from negative to positive.
Since the state continued from the $n$th Efimov state has the $n$th lowest $\mathrm{Re}[E]$ over most, though not all, of the range of $s$, we hereafter call $n=1,2,3,4$ the ground, first, second, and third excited states.
Because \equref{hamiltonian_1} is scale invariant apart from the length scale $r_\mathrm{min}$, the trajectories in
\figref{complex_pot_localized_state_energy_1}(a) are self-similar under rescaling of $r_\mathrm{min}$.
This is a complex-plane manifestation of the quantum anomaly underlying the Efimov effect~\cite{PhysRevD.68.125013,Braaten2006259}, where $r_\mathrm{min}$ breaks the continuous scale invariance down to a discrete one.
The rotation is faster for larger $n$, as governed mainly by the $e^{-2\pi i n/s}$ factor in \equref{energy_bound_state_analytical}; indeed, as shown in \figref{complex_pot_localized_state_energy_1}(b), $\arg E$ increases faster for larger $n$.
This suggests that the higher excited states with smaller $|E|$ are more sensitive to the dissipation described by $\mathrm{Im}[s^{2}]$. 
As the strength of $\mathrm{Im}[s^{2}]$ is further increased, the states disappear successively from the highest excited state, because the energy must
satisfy $\pi \le \arg E < 2\pi$
($\pi/2 \le \arg k < \pi$).
This condition follows from $\mathrm{Im}[E]\le 0$, which results from $\mathrm{Im}[s^{2}]\le 0$, together with the fact that the states should lie on the first Riemann sheet since they are continuously connected to the Efimov bound states.

This successive disappearance can also be understood from \equref{energy_bound_state_analytical}. The requirement $\pi \le \arg E < 2\pi$ dictates the condition on $n$ as 
\begin{equation}
    \label{eq:n_condition}
    m \leq \frac{1}{2\pi}\mathrm{Im}\left[\frac{1}{s}
     \left\{\ln\left(\frac{-\Gamma(s)}{\Gamma(-s)}\right) - 2\pi i n\right\}\right]
    < m+\frac{1}{2},
\end{equation}
where the integer $m$ is taken to be zero for the localized states in \figref{complex_pot_localized_state_energy_1}, since they connect naturally to the Hermitian limit with this choice. The number of localized states is thus finite and changes as $s$ is varied, with the condition violated first for larger $n$, in accordance with the successive disappearance from the higher excited states in \figref{complex_pot_localized_state_energy_1} (a). 
A similar counterclockwise rotation and disappearance of the complex energies have recently been reported for a two-body system with a complex zero-range potential, including the scale-invariant case in two dimensions~\cite{TajimanonHermite2026}.

\begin{figure}[!t]
    \centering
    \includegraphics[width=1.0\linewidth]{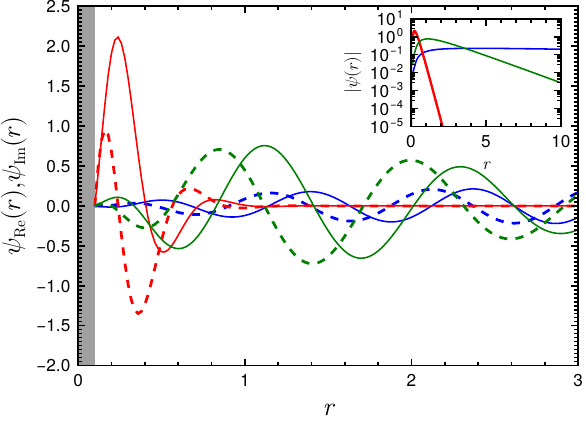} 
\caption{Eigenfunctions obtained numerically from \equref{Discretization_Hamiltonian}, corresponding to the filled circles in \figref{complex_pot_localized_state_energy_1} at $|s| \approx 3.163$ and $\arg s \approx 0.669 \pi$ (i.e., $s^{2} - 1/4 \approx -5.107 - 8.745 i$). The red and green curves are the localized ground and first excited states, and the blue curve is the continuum state at $E \approx 39.499 - 0.467 i$. Main panel: real (solid) and imaginary (dashed) parts. Grey area: region $r < r_\mathrm{min}$ excluded by the boundary condition. Inset: absolute value $|\psi(r)|$.}
\label{fig:complex_pot_localized_state_wavefunc_1}
\end{figure}

We can also obtain the analytical expression of the energies of the complex continuum states from \equref{wavefunc_hankel_1}. Indeed, by imposing the boundary conditions
$\psi(r=r_\mathrm{min}) = \psi(r=r_\mathrm{max}) = 0$, we obtain $\displaystyle H_s^{(1)}(kr_\mathrm{min})/ H_s^{(2)}(kr_\mathrm{min})= H_s^{(1)}(kr_\mathrm{max})/H_s^{(2)}(kr_\mathrm{max})$. As in the case of the localized states, we assume that the imaginary part is
not too large, $-1<\mathrm{Re}[s]<1$, and take the branch of $s$ such that
$\mathrm{Re}[s]<0$. Using the asymptotic forms of the Hankel functions at
short distance, $|kr_\mathrm{min}|\to 0$, and at large distance,
$|kr_\mathrm{max}|\to\infty$ (see Appendix~\ref{app:Analytic_derivation}), we find $\displaystyle  e^{2i\left(kr_\mathrm{max} - \pi/4 - \pi s/2 \right)} = -e^{-2is\pi}$, from which we obtain
\begin{equation}
    \label{eq:wavenumber_continuum_state_analytical_2}
    k = \frac{\pi}{4r_\mathrm{max}}\left(4n -2s +3 \right),
\end{equation}
\begin{equation}
    \label{eq:energy_continuum_state_analytical_2}
    E = \frac{1}{2\mu}\left(\frac{\pi}{4r_\mathrm{max}}\right)^2\left(4n - 2s + 3\right)^2.
\end{equation}
The energy, shown by the solid curve in \figref{complex_pot_localized_state_energy_1}(c), acquires a small negative imaginary part for complex $s$. Note that the continuum states described by \equref{energy_continuum_state_analytical_2} lie on the second Riemann sheet of the complex energy plane, because they are continuously connected to the real continuum states at $s^{2}<0$ as $\mathrm{Im}[s^{2}]\to 0$. 
This is consistent with \equref{wavenumber_continuum_state_analytical_2}: 
for $\mathrm{Re}[s]<0$ and $\mathrm{Im}[s]> 0$, it gives $3\pi/2<\arg k< 2\pi$, which lies in the second Riemann sheet. 
Since the localized states described by \equref{energy_bound_state_analytical} lie on a different Riemann sheet, no avoided crossing or exceptional point (discussed below) is expected between these two kinds of states, even when they happen to take the same complex energy value.

We can also obtain essentially the same results using the WKB approximation. As described in detail in Appendix~\ref{app:WKB}, we obtain the energy of the localized state as
\begin{equation}
    \label{eq:energy_bound_state_WKB_1}
    E = \frac{1}{2\mu}\left(\frac{2}{r_\mathrm{min}}\right)^2 \exp\left[\frac{\pi i}{2s} -2 +2\ln \left(s\right)\right]\exp\left[- \frac{2\pi i}{s}n\right].
\end{equation}
This result is shown by the dashed curves in \figref{complex_pot_localized_state_energy_1} (a)(b), which agree excellently with the analytical result in \equref{energy_bound_state_analytical} (solid). In particular, they differ only slightly in the prefactor, whereas the $e^{-2\pi i n/s}$ term predominantly describing the rotation in the complex plane is reproduced. The continuum state energy obtained by
the WKB approximation exactly agrees with \equref{energy_continuum_state_analytical_2} (see Appendix~\ref{app:WKB}).

These analytical and WKB results can also be reproduced by solving \equref{hamiltonian_1} numerically; we discretize it by the three-point finite difference method between the boundaries $r_\mathrm{min}$ and $r_\mathrm{max}$ with a uniform mesh size $h=(r_\mathrm{max}-r_\mathrm{min})/(N+1)$, and numerically compute the eigenvalues and eigenvectors of the non-Hermitian tridiagonal matrix
\begin{equation}
    \label{eq:Discretization_Hamiltonian}
    2\mu \hat{H} =
    \begin{pmatrix}
        d_{1} & -t &  &  & 0 \\
        -t & d_{2} & -t &  &  \\
         & \ddots & \ddots & \ddots &  \\
         &  & -t & d_{N-1} & -t \\
        0 &  &  & -t & d_{N}
    \end{pmatrix},
\end{equation}
where $t \equiv 1/h^{2}$ and $d_{j} \equiv 2/h^{2} + (s^{2}-1/4)/r_{j}^{2}$ with $r_j=r_\mathrm{min} + hj$ ($j=1,2,..,N$). The numerical results agree excellently with the analytical and WKB curves, both for the localized states in \figref{complex_pot_localized_state_energy_1}(a) and (b) (circles), and for the continuum states in (c) (dotted curve). The slight disagreement, seen for the ground state as well as for the continuum states, is attributed in both cases to the breakdown of the low-energy approximation used in deriving Eqs.~(\ref{eq:energy_bound_state_analytical}) and (\ref{eq:energy_continuum_state_analytical_2}).

In \figref{complex_pot_localized_state_wavefunc_1}, we show the real part (solid) and imaginary part (dashed) of the numerically obtained eigenfunctions of the localized states (red, green) and the continuum state (blue). The former are concentrated around the distance $r\sim r_{\mathrm{min}}$, while the latter extends to large distance. This is consistent with our results in Eqs.~(\ref{eq:energy_bound_state_analytical}) and (\ref{eq:energy_continuum_state_analytical_2}): the energies of the localized and continuum states are primarily controlled by $r_{\mathrm{min}}$ and $r_{\mathrm{max}}$, respectively. The exponential decay of the localized state (inset) is mainly governed by the imaginary part of the wave number $k=\sqrt{2\mu E}$ since $H^{(1)}_s(kr) \sim e^{i k r} \propto e^{- \mathrm{Im}[k] r}$.

\begin{figure}[!t]
    \centering
    \includegraphics[width=1.0\linewidth]{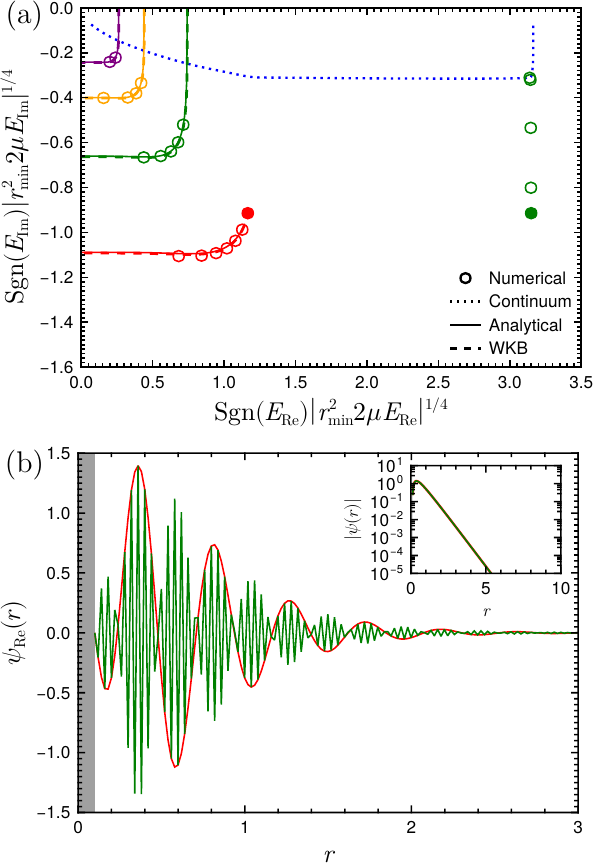} 
\caption{(a) Energies of the localized states near the end point of
\figref{complex_pot_localized_state_energy_1}(a), where the potential becomes purely imaginary, $s^{2} - 1/4 = -10.0i$, obtained with the same parameters and symbols. The blue dotted curve is the continuum state for the purely imaginary potential. (b) Eigenfunctions of the localized ground (red) and first excited (green) states for the purely imaginary potential, corresponding to the terminating filled circles in (a), respectively. Main panel: real part. Inset: absolute value $|\psi(r)|$.}
\label{fig:to_pure_imag_and_localized_wavefunc}
\end{figure}

As the complex potential strength is varied toward a purely imaginary potential $s^2-1/4 = - i \beta$ ($\beta >0 $) in \figref{complex_pot_localized_state_energy_1}~(a), corresponding to the two heavy polarons immersed in a finite-temperature quantum medium~\cite{PhysRevA.110.033304}, the states continued from the excited Efimov states successively disappear, and only the lowest few remain localized. 
For the value of $\beta$ used in \figref{complex_pot_localized_state_energy_1}~(a), the analytical solution \equref{energy_bound_state_analytical} yields a single localized state, while the number of such states increases with $\beta$ (see Appendix~\ref{app:beta_crit_eval}).
As shown in the magnified plot in \figref{to_pure_imag_and_localized_wavefunc}~(a), the ground Efimov state (red) connects smoothly to a localized eigenstate in the fourth quadrant of the complex energy plane under a purely imaginary potential (red filled circle). 
This localized state arises solely from the imaginary potential, and the mechanism of its localization is therefore distinct from that of the states bound by the attractive real part of the potential. 
The emergence of this localized
state demonstrates that the Efimov state at $s^{2}<0$, a paradigmatic example of quantum three-body physics, is analytically continued in the complex plane into a localized bipolaron state arising from environment-induced dissipation. The complex $1/r^{2}$ potential problem thus bridges few-body Hermitian physics and many-body non-Hermitian physics.

Although the analytical expression in \equref{energy_bound_state_analytical} is derived under the condition $-1<\mathrm{Re}[s]<1$, which is not strictly satisfied for the purely imaginary potential, it still agrees reasonably well with the numerical results in \figref{to_pure_imag_and_localized_wavefunc}~(a). 
Notably, there is one more localized state (green filled circle) that is not captured by the analytical solution but appears in the finite-difference numerical calculation.
This state has a value of $\mathrm{Im}[2\mu E] \sim 1/r_{\mathrm{min}}^2$ similar to \equref{energy_bound_state_analytical}, but a significantly larger real part $\mathrm{Re}[2\mu E] \sim 4/h^2$. 
Since the analytical solution describes the continuum limit $h\to 0$, this state is specific to the discretized Hamiltonian, with an energy scale set by the mesh size $h$.
In \figref{to_pure_imag_and_localized_wavefunc}~(a), the state continued from the first excited Efimov state (green circle) disappears at $\mathrm{Re}[2\mu E] \sim 1/r_{\mathrm{min}}^2$, and subsequently
reappears at $\mathrm{Re}[2\mu E] \sim 4/h^2$, becoming this second localized state as $s^2 -1/4$ is varied toward the purely imaginary value. The disappearance and reappearance do not necessarily imply that the two states are unrelated. For a different choice of how the real and imaginary parts of $s^{2}-1/4$ are varied, the first excited state is found to evolve smoothly into this second localized state, indicating that the disappearance and reappearance depend on the path taken in the complex parameter space.

The first (red) and second (green) localized states also have markedly different eigenfunctions. As shown in \figref{to_pure_imag_and_localized_wavefunc}~(b), the oscillation wavelength of the first localized state (red) is $\sim r_{\mathrm{min}}$, whereas the second localized state (green) oscillates much faster $\sim h$. This is consistent with $\mathrm{Re}[2\mu E]\sim 1/r_{\mathrm{min}}^2$ and  $\mathrm{Re}[2\mu E]\sim 4/h^2$. 
On the other hand, their envelopes and the exponential decay rates at large distance (inset of \figref{to_pure_imag_and_localized_wavefunc}~(b)) are the same. This agreement originates from a symmetry
of the discretized Hamiltonian, as discussed below.

\subsection{Symmetry and exceptional points of the discretized Hamiltonian}
The second localized state (green filled circle) in \figref{to_pure_imag_and_localized_wavefunc}~(a) does not appear in the continuum Schr\"odinger equation~(\ref{eq:hamiltonian_1}) and hence has no direct physical consequence for the bipolaron physics; such a state is sometimes referred to as a spurious state in the literature~\cite{LAKOBA2013117,PhysRevC.106.L051303}. It nevertheless reflects a fundamental symmetry of the discretized Hamiltonian for the purely imaginary potential. To see this point, we introduce an $N \times N$ matrix $\hat{C} = 2\mu\hat{H} - 2/h^2 \hat{I}$, where $\hat{I}$ is the identity matrix. This shifted Hamiltonian $\hat{C}$ has the same eigenvectors as $2\mu\hat{H}$ with the eigenvalues shifted uniformly by $2/h^2$. The matrix $\hat{C}$ possesses the symmetry 
\begin{equation}
\left(\mathcal{S}\mathcal{T}\right)\hat{C}\left(\mathcal{S}\mathcal{T}\right)^{-1} = -\hat{C},
\label{eq:STsym_Cmat}
\end{equation}
if and only if the potential is purely imaginary. Here, $ \mathcal{S}=\mathrm{diag}\left[\left(-1\right)^j\right]$ is a diagonal unitary matrix, and $\mathcal{T}$ is the time-reversal operator satisfying $\mathcal{T} i \mathcal{T}^{-1} = -i$. 
Since $\mathcal{ST}$ is antiunitary,
Eq.~(\ref{eq:STsym_Cmat}) implies that $i\hat{C}$ commutes
with $\mathcal{ST}$:
$\left(\mathcal{S}\mathcal{T}\right) i\hat{C}
\left(\mathcal{S}\mathcal{T}\right)^{-1} = i \hat{C}$.
As shown in Appendix~\ref{app:proof_st},
the eigenvalues of $i\hat{C}$ are either real
or appear as complex conjugate pairs, in the same manner
as for $\mathcal{PT}$-symmetric
Hamiltonians~\cite{PhysRevLett.80.5243}.
 It follows that the energy eigenvalues of the discretized Hamiltonian $\hat{H}$ either lie on the line $\mathrm{Re}[2\mu E]=2/h^2$, or form pairs symmetric about this line when the $\mathcal{ST}$ symmetry is spontaneously broken.

Equation~(\ref{eq:STsym_Cmat}) can be written as
$\mathcal{S}\hat{C}^{*}\mathcal{S}^{-1}=-\hat{C}$,
which has the same algebraic form as the
$\mathrm{PHS}^{\dagger}$ relation in the symmetry
classification of non-Hermitian
Hamiltonians~\cite{kawabata19}.
Since $\hat{C}^{\dagger}=\hat{C}^{*}$, this relation can equivalently be 
written as $\mathcal{S}\hat{C}^{\dagger}\mathcal{S}^{-1}=-\hat{C}$,
which has the same algebraic form as the chiral symmetry
in the same classification.
A closely related algebraic structure appears in a non-Hermitian Su--Schrieffer--Heeger model with alternating gain and loss, whose symmetry leads to the spectral pairing $(E, -E^*)$ and a symmetry-breaking transition through an exceptional point~\cite{PhysRevB.98.094307,PhysRevB.99.121101}. 
We emphasize, however, that $\mathcal{S}$ originates from
the finite-difference discretization and does not represent a physical
particle-hole or chiral symmetry of the continuum Hamiltonian
in Eq.~(\ref{eq:hamiltonian_1}), which is consistent with the spurious nature of the second
localized state.

\begin{figure}[!t]
    \centering
    \includegraphics[width=1.0\linewidth]{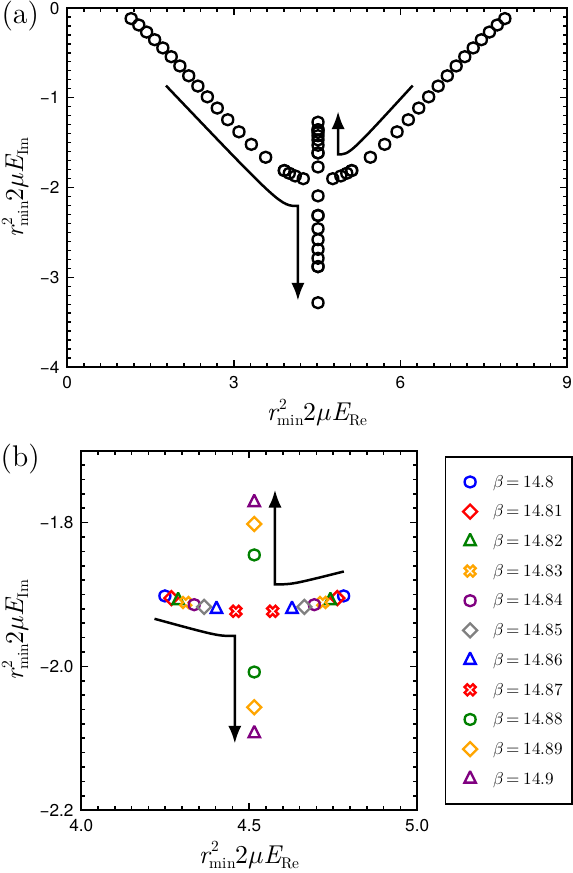} 
\caption{Trajectories of the complex energy eigenvalues of the two localized states for a purely imaginary potential $s^2 -1/4 = -i \beta$. $\beta$ is increased (a) from $7.0$ to $16.0$ in steps of $0.5$ (from $14.5$ to $15.5$ near the exceptional point, in steps of $0.1$),
and (b) from $14.8$ to $14.9$ in steps of $0.01$  ($r_\mathrm{min}=0.1$, $r_\mathrm{max}=100$, $N=1500$). Arrows indicate the direction of increasing $\beta$.}
\label{fig:pure_imag_pot_localized_state_energy_1}
\end{figure}

\begin{figure}[!t]
    \centering
    \includegraphics[width=1.0\linewidth]{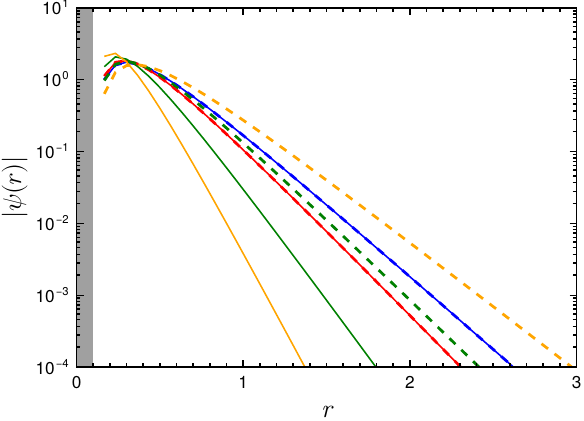} 
\caption{Eigenfunctions of the localized states for various strengths $\beta$ of purely imaginary potential ($r_\mathrm{min}=0.1$, $r_\mathrm{max}=100$, $N=1500$). For $\beta < \beta^{(\mathrm{ep})}$, $\beta=13.0$ (blue) and  $14.0$ (red), the two localized states (solid and dashed) have the same absolute values, whereas for $\beta > \beta^{(\mathrm{ep})}$, $\beta=15.0$ (green) and $16.0$ (yellow), they turn into strongly (solid) and weakly (dashed) localized states.}
\label{fig:pure_imag_pot_localized_state_wavefunc_1}
\end{figure}

To examine the consequences of the $\mathcal{ST}$ symmetry of \equref{STsym_Cmat}, we vary the strength of the purely imaginary potential $s^2 -1/4 = -i \beta$ and show the behavior of the pair of localized states in \figref{pure_imag_pot_localized_state_energy_1}. The two localized states appear symmetrically about the line $\mathrm{Re}[2\mu E]=2/h^2$ for $\beta < \beta^{(\mathrm{ep})}$;
the symmetry of the discretized non-Hermitian Hamiltonian thus makes the second localized state an inevitable
consequence of the first. At the critical value $\beta = \beta^{(\mathrm{ep})}$ (see Appendix~\ref{app:beta_crit_eval} for its detailed evaluation), the two states collide on the line $\mathrm{Re}[2\mu E]=2/h^2$, and then move in opposite directions along it. The corresponding change of the eigenvectors is shown in \figref{pure_imag_pot_localized_state_wavefunc_1}. 
For $\beta < \beta^{(\mathrm{ep})}$, the two states share the same localization length, whereas for $\beta > \beta^{(\mathrm{ep})}$, they separate into more strongly (solid) and more weakly (dashed) localized states. The two regimes correspond to the spontaneously broken and unbroken phases of the $\mathcal{ST}$ symmetry. 
For $\beta < \beta^{(\mathrm{ep})}$, the two energies are located symmetrically about the line $\mathrm{Re}[2\mu E]=2/h^{2}$,
and $\mathcal{ST}$ maps one state onto the other as $\mathcal{ST}\psi_{1}=\psi_{2}$ with the phases of the eigenstates chosen appropriately (see Appendix~\ref{app:proof_st}). This yields $\psi_{2}(r_{j})=(-1)^{j}\psi_{1}^{*}(r_{j})$: the two states have the same localization length,
while their relative sign alternates from site to site, as demonstrated in \figref{to_pure_imag_and_localized_wavefunc}~(b). 
For $\beta > \beta^{(\mathrm{ep})}$, on the other hand, both energies lie on the line, and hence $\mathcal{ST}\psi_{n}=\psi_{n}$, allowing the two states to have different localization lengths. 
We have numerically confirmed that the two eigenfunctions coincide at $\beta = \beta^{(\mathrm{ep})}$, which identifies this point as an exceptional point of the non-Hermitian matrix.
This symmetry-enabled exceptional point is of a different origin from those recently reported in Ref.~\cite{pf6y-lxzp}, which occur between a resonance and the rotated continuum; in our case, the continuum states lie on a different Riemann sheet and cannot produce an exceptional point with the localized states.

\begin{figure}[!t]
    \centering
    \includegraphics[width=1.0\linewidth]{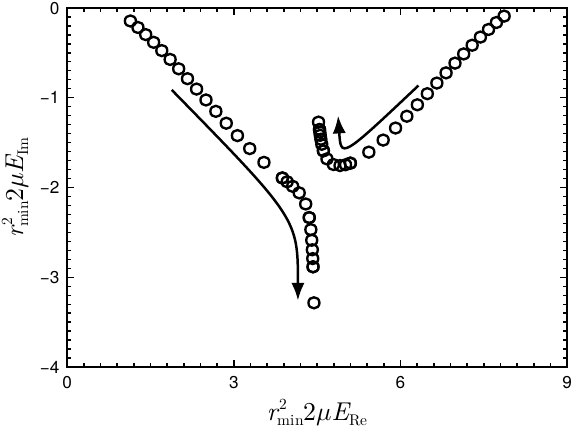} 
\caption{Trajectories of the complex energy eigenvalues of the two localized states
for a non-purely-imaginary potential $s^2 -1/4 = -0.1 -i \beta$ as $\beta$ is increased from $7.0$ to $16.0$ in steps of $0.5$ (from $14.5$ to $15.5$, in steps of $0.1$). Arrows indicate the direction of increasing $\beta$ ($r_\mathrm{min}=0.1$, $r_\mathrm{max}=100$, $N=1500$).}
\label{fig:non_pure_imag_pot_localized_state_energy_1}
\end{figure}

The exceptional point is found only for a purely imaginary potential. In \figref{non_pure_imag_pot_localized_state_energy_1}, we show the energies of the pair of localized states for a potential that is not purely imaginary. As the magnitude $\beta$ of the imaginary part is increased with a finite real part, $s^2 - 1/4 = -0.1 - i\beta$, the two states move in the complex plane, but never collide. The energies are also no longer symmetric about $\mathrm{Re}[2\mu E]=2/h^2$, demonstrating that the symmetry and the exceptional point are specific to the purely imaginary potential. Notably, the symmetry in \equref{STsym_Cmat} is not unique to the $1/r^2$ potential studied in this work; it holds for any potential shape if the potential strength is purely imaginary. The pair of localized states therefore appear generically in the discretized non-Hermitian Schr\"odinger equation with a purely imaginary potential, and can form an exceptional point as the imaginary potential strength is varied.

%************************************************************************
\section{\label{sec:Concl}Conclusion}

We have studied the non-Hermitian Schr\"odinger equation with a complex $1/r^{2}$ potential, which describes the Efimov states for a real attractive potential and two heavy polarons in a degenerate many-body medium for a purely imaginary potential. We have obtained analytical solutions for the localized and continuum states in terms of Hankel functions of complex index, in agreement with numerical solutions and the WKB approximation. As the potential is varied from a real $1/r^{2}$ attraction to a purely imaginary one, the ground Efimov state continuously evolves into a localized bipolaron state arising solely from the medium-induced imaginary interaction. In a discretized formulation, this state is accompanied by a second localized state that oscillates on the scale of the mesh spacing, and the two states form an exceptional point when the potential is purely imaginary. We have shown that this exceptional point originates from a symmetry of the Hamiltonian discretized on a uniform mesh, valid for any purely imaginary potential.

Although the partner localized state and the exceptional point have no counterpart in the continuum limit, the bipolaron state itself can be realized in highly mass-imbalanced cold-atom mixtures by immersing heavy impurities in a Bose--Einstein condensate or a degenerate Fermi gas of light atoms. The real and imaginary parts of its energy would be accessible from RF spectroscopy~\cite{PhysRevLett.102.230402,Kohstall_2012,Koschorreck_2012,PhysRevLett.117.055301,PhysRevLett.117.055302,PhysRevLett.118.083602,PhysRevLett.122.093401,Yan_2020} or from the damped oscillations in Ramsey spectroscopy~\cite{PhysRevB.94.184303,Cetina_2016,PhysRevResearch.7.L022002}, by properly subtracting the single-polaron contributions. 
A quantitative comparison with such measurements requires further theoretical studies including the real part of the medium-induced interaction, which is not negligible at shorter distances~\cite{PhysRev.96.99,10.1143/PTP.16.45,PhysRev.106.893,PhysRevLett.129.233401,DeSalvo:2019juk,Baroni:2023ejp,PhysRevLett.131.083003,w7rd-2qpv,muir2022interactions,PhysRevX.13.031036}, as well as medium effects beyond the non-Hermitian description.
Such experimental and theoretical studies would elucidate the open quantum nature of polarons, and thereby advance a unified understanding of few-body and many-body physics.

%%%%%%%%%%%%%%%%% 
\begin{acknowledgments}
 We thank Yukinao Akamatsu, Yasuyuki Hatsuda, Masaru Hongo, Tetsuo Hyodo, Masaya Nakagawa, Shoya Ogawa, Hiroyuki Tajima, and Shion Yamashika for fruitful discussions. SE acknowledges support from Matsuo Foundation, JGC-Saneyoshi Scholarship Foundation, and JSPS KAKENHI Grant Numbers JP23H01174, and JP25K00217. KY acknowledges support from JSPS Program for Forming Japan's Peak Research Universities (J-PEAKS) Grant No.\ JPJS00420230008, JSPS KAKENHI Grant No. JP25K17327, Hirose Foundation, Fujikura Foundation, Toyota Riken Scholar Program, and Support Center for Advanced Telecommunications Technology Research. We used Claude Opus 5 and 5.5 (Anthropic) to polish the English, and Claude Fable 5 (Anthropic) to suggest a candidate symmetry of the matrix in \equref{STsym_Cmat}, which we then proved analytically. We have carefully checked all the AI-assisted outputs.
\end{acknowledgments}

%%%%%%%%%%%%%%%%% 
\appendix

\section{\label{app:Analytic_derivation}Derivation of the analytical results}
In this Appendix, we derive Eqs.~(\ref{eq:energy_bound_state_analytical}) and (\ref{eq:energy_continuum_state_analytical_2}) from the solution of the
Schr\"odinger equation (\ref{eq:hamiltonian_1}), which is expressed as a
linear combination of the Hankel functions of complex
index as in \equref{wavefunc_hankel_1}. For the localized states, we
first use the asymptotic forms of the Hankel functions at $|kr|\gg 1$,
\begin{equation}
    H^{(1,2)}_s(kr)\rightarrow\sqrt{\frac{2}{\pi kr}}\exp\left[\pm i\left(kr-\frac{\pi}{4}-\frac{\pi}{2}s\right)\right].
\end{equation}
Since the localized states lie on the first Riemann sheet, i.e., $\mathrm{Im}[k]>0$, $H^{(1)}_s(kr)$ decays exponentially at large distances, whereas $H^{(2)}_s(kr)$ grows exponentially. 
Because the localized states must satisfy $|\psi(r)| \rightarrow 0$ at large distances, $c_{2}=0$ follows for $|kr_{\mathrm{max}}|\gg 1$, and hence $\psi(r) = c_1\sqrt{r}H_s^{(1)}(kr)$. The Hankel functions are further expanded as
\begin{equation}
    \begin{split}
    H_s^{(1,2)}(kr) ={}& \frac{\pm ie^{\mp is\pi}}{\sin(s\pi)}
    \sum_{m=0}^{\infty}\frac{\left(-1\right)^m}
    {m!\Gamma\left(s+m+1\right)}\left(\frac{kr}{2}\right)^{2m+s}\\
    &\mp \frac{i}{\sin(s\pi)}\sum_{m=0}^{\infty}
    \frac{\left(-1\right)^m}{m!\Gamma\left(-s+m+1\right)}
    \left(\frac{kr}{2}\right)^{2m-s}.
    \end{split}
\end{equation}
At low energies $|kr|\to 0$, the higher-order terms of the first sum are smaller than the $m=0$ term of the second sum if $\mathrm{Re}[s]> -1$.
Similarly, the higher-order terms of the second sum are smaller than the $m=0$ term of the first sum if $\mathrm{Re}[s]< 1$. 
Therefore, for $-1<\mathrm{Re}[s]<1$, the low-energy Hankel functions take the asymptotic form
\begin{equation}
    \label{eq:HankelAsymp_scond}
    \begin{split}
    H_s^{(1,2)}(kr) \to \frac{\pm i}{\sin(s\pi)}\bigg[&
    \frac{e^{\mp is\pi}}{\Gamma(1+s)}\left(\frac{kr}{2}\right)^{s}\\
    &- \frac{1}{\Gamma(1-s)}\left(\frac{kr}{2}\right)^{-s}\bigg].
    \end{split}
\end{equation}
Using the Gamma function formula $\Gamma(s)\Gamma(1-s)=\pi / \sin(\pi s)$, this expression is rewritten as
\begin{equation}
\label{eq:HankelAsymp_scond_2}
    H_s^{(1,2)}(kr) = \mp\frac{i}{\pi}\left[e^{\mp is\pi}\Gamma(-s)\left(\frac{kr}{2}\right)^{s} + \Gamma(s)\left(\frac{kr}{2}\right)^{-s}\right].
\end{equation}
When the low-energy condition $|kr_\mathrm{min}|\ll 1$ is satisfied, $\psi (r=r_\mathrm{min}) = 0$ leads to
\begin{equation}
    \left(\frac{kr_\mathrm{min}}{2i}\right)^{2s} = \frac{-\Gamma(s)}{\Gamma(-s)},
\end{equation}
which gives, with an arbitrary integer $n$, 
\begin{equation}
    \frac{kr_\mathrm{min}}{2i} = \exp\left[\frac{1}{2s}\ln \left(\frac{-\Gamma(s)}{\Gamma(-s)}\right) - \frac{2n\pi i}{2s}\right],
\end{equation}
from which \equref{energy_bound_state_analytical} follows.

For the continuum states, on the other hand, the boundary
condition $\psi(r=r_\mathrm{max}) = 0$ gives
\begin{equation}
    \psi(r) = c\sqrt{r}\left[ H_s^{(1)}(kr) -  \frac{H_s^{(1)}(kr_\mathrm{max})}{H_s^{(2)}(kr_\mathrm{max})}H_s^{(2)}(kr) \right].
\end{equation}
Imposing the other boundary condition $\psi(r=r_\mathrm{min}) = 0$ then yields
\begin{equation}
    \label{eq:Hankel_gen_trans_eq}\frac{H_s^{(1)}(kr_\mathrm{min})}{H_s^{(2)}(kr_\mathrm{min})} = \frac{H_s^{(1)}(kr_\mathrm{max})}{H_s^{(2)}(kr_\mathrm{max})}.
\end{equation}
For the left-hand side, \equref{HankelAsymp_scond_2} gives, under the
condition $-1<\mathrm{Re}[s]<1$,
\begin{equation}
    \frac{H_s^{(1)}(kr_\mathrm{min})}{H_s^{(2)}(kr_\mathrm{min})} = - \frac{e^{-is\pi}\Gamma(-s)\left( \frac{kr_\mathrm{min}}{2} \right)^{s} + \Gamma(s)\left( \frac{kr_\mathrm{min}}{2} \right)^{-s}}{e^{is\pi}\Gamma(-s)\left( \frac{kr_\mathrm{min}}{2} \right)^{s} + \Gamma(s)\left( \frac{kr_\mathrm{min}}{2} \right)^{-s}}.
\end{equation}
Since we consider $\mathrm{Re}[s]<0$ in this paper, the asymptotic form
\begin{equation}
    \frac{H_s^{(1)}(kr_\mathrm{min})}{H_s^{(2)}(kr_\mathrm{min})} \to -e^{-2is\pi}
\end{equation}
is obtained. For the right-hand side of \equref{Hankel_gen_trans_eq}, we have
\begin{equation}
    \frac{H_s^{(1)}(kr_\mathrm{max})}{H_s^{(2)}(kr_\mathrm{max})} \rightarrow e^{2i(k r_\mathrm{max} -\pi/4-\pi s/2)}.
\end{equation}
We thus obtain the equation determining the energies as
\begin{equation}
    -e^{-2is\pi} = e^{2i(k r_\mathrm{max} -\pi/4-\pi s/2)}.
\end{equation}
Taking the logarithm of both sides, we obtain, with $n$ being an
arbitrary integer, 
\begin{equation}
%    \label{eq:wavenumber_continuum_state_analytical_2}
    k = \frac{\pi}{4r_\mathrm{max}}\left(4n -2s +3 \right),
\end{equation}
from which \equref{energy_continuum_state_analytical_2} follows.

%%%%%%%%%%%%%%%%%%%%%%%%%%%%%%%%%%%%%%%%%%%%%%%%%%%%%%%%%%%%%%%%%%
\section{\label{app:WKB}Derivation of the WKB energies}
To derive the WKB solutions, we use the classical turning point
\begin{equation}
    r_{t} =\sqrt{\frac{s^2}{2\mu E}},
\end{equation}
which is obtained by taking into account the Langer's correction~\cite{PhysRev.51.669}. The complex energies of the localized states and continuum states can be derived from the WKB quantization condition with this turning point.

For the localized states, the Bohr--Sommerfeld quantization condition is imposed between $r_\mathrm{min}$ and $r_t$, where the phase shifts acquired at the two turning points require careful treatment~\cite{maslov_fedoriuk_1981,PhysRevA.54.1136}. 
Specifically, since a hard-wall boundary condition is imposed at $r_{\mathrm{min}}$, the phase shift there is $\pi$, whereas at $r_{t}$ the phase shift is $\pi/2$ in the low-energy limit $|r_{t}| \rightarrow \infty$. The two phase shifts add up to a total Maslov index of $3$, so that the quantization condition reads
\begin{equation}
   \label{eq:BS_localized} \left(n + \frac{3}{4}\right)\pi = \int_{r_\mathrm{min}}^{r_t}dx\sqrt{2\mu E - \frac{s^2}{x^2}}.
\end{equation}
Since $r_t$ is complex, both the branch of the square root
and the integration contour in \equref{BS_localized} must be chosen carefully.
For the square root in the integrand, the branch cut is taken slightly below the positive real axis, because the poles of the localized states must lie in the first Riemann sheet.
For the integration contour, we continuously deform the real-axis contour of the Hermitian limit, where $r_t$ is real, without crossing the singularities of the integrand.
Since the integrand is analytic in the region swept by this deformation, \equref{BS_localized} does not depend on the details of the contour.
Changing the integration variable in \equref{BS_localized}, we obtain
\begin{equation}
    \left(n + \frac{3}{4}\right)\pi = \int^{0}_{\xi_{\mathrm{min}}}du\frac{u^2}{u^2 + s^2},
\end{equation}
where $\xi_{\mathrm{min}} = \sqrt{2\mu E r_\mathrm{min}^{2} - s^{2}}$. Performing the integral yields
\begin{equation}
    \left(n + \frac{3}{4}\right)\pi =  - \xi_{\mathrm{min}} - \frac{is}{2}\ln \left(-\frac{\xi_{\mathrm{min}} - is}{\xi_{\mathrm{min}} + is}\right).
\end{equation}
A low-energy expansion for $\sqrt{2\mu |E|} r_\mathrm{min} \ll |s|$ then gives
\begin{equation}
    \left(n + \frac{3}{4}\right)\pi = is - \frac{is}{2}\ln\left(\frac{4s^2}{2\mu E{r_\mathrm{min}}^2}\right),
\end{equation}
where we have used $\sqrt{-s^2}= - is$, consistent with the branch cut of the square root. Solving this for $E$ yields
\begin{equation}
    E = \frac{1}{2\mu} \left(\frac{2}{r_\mathrm{min}}\right)^2 \exp\left[ - \left(2n+\frac{3}{2}\right)\frac{\pi i}{s} -2 +2\ln \left(s\right)\right],
\end{equation}
from which \equref{energy_bound_state_WKB_1} is obtained by replacing the arbitrary integer $n$ with $n-1$. The requirement $\pi \le \arg E < 2\pi$ leads to the following condition on $n$ for the localized state:
\begin{equation}
    \label{eq:n_condition_WKB}
    m-\frac{1}{2} \leq \frac{1}{2\pi}\mathrm{Im}\left[ - \left(2n+\frac{3}{2}\right)\frac{\pi i}{s} +2\ln \left(s\right)\right]
    < m,
\end{equation}
where $m$ is an integer.

For the continuum states, the Bohr--Sommerfeld quantization condition is imposed between $r_t$ and $r_\mathrm{max}$. Similarly to the localized states, the $\pi$ phase shift at the hard-wall boundary $r_\mathrm{max}$ and $\pi/2$ phase shift at $r_t$ add up to the quantization condition with a Maslov index~\cite{maslov_fedoriuk_1981,PhysRevA.54.1136} of 3:
\begin{equation}
    \left(n + \frac{3}{4}\right)\pi = \int_{r_t}^{r_\mathrm{max}}dx\sqrt{2\mu E - \frac{s^2}{x^2}}.
\end{equation}
In contrast to the localized states, the branch cut of the square root must be taken on the negative real axis because the continuum states lie on the second Riemann sheet. Performing the integral in the same manner as above, we obtain 
\begin{equation}
    \left(n + \frac{3}{4}\right)\pi = \xi_{\mathrm{max}} + \frac{is}{2}\ln\left(-\frac{\xi_{\mathrm{max}} - is}{\xi_{\mathrm{max}} + is}\right),
\end{equation}
where $\xi_{\mathrm{max}} =  \sqrt{2\mu E r_\mathrm{max}^{2} - s^{2}}$. Assuming $|s| \ll \sqrt{2\mu |E|} r_\mathrm{max} $ and taking the branch of $\mathrm{ln} (-1)= - i\pi$, we obtain 
\begin{equation}
    \left(n + \frac{3}{4}\right)\pi = \sqrt{2\mu E}r_\mathrm{max} + \frac{s\pi}{2},
\end{equation}
from which it follows that 
\begin{equation}
    E = \frac{1}{2\mu}\left(\frac{\pi}{4r_\mathrm{max}}\right)^2\left(4n - 2s + 3\right)^2.
\end{equation}
Note that this WKB result coincides with the analytical one in \equref{energy_continuum_state_analytical_2}.

\begin{figure}[!t]
    \centering
    \includegraphics[width=1.0\linewidth]{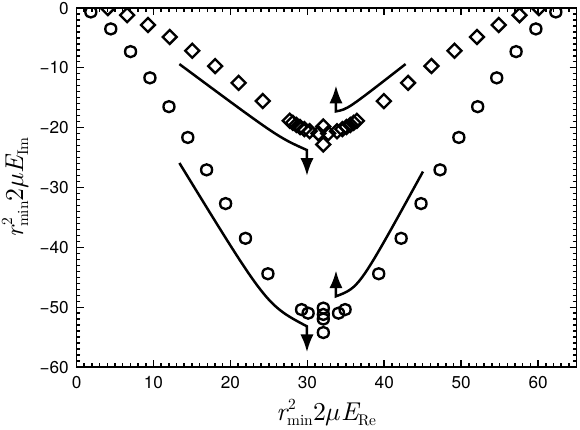} 
\caption{Trajectories of the complex energy eigenvalues of the four localized states for a purely imaginary potential $s^2 -1/4 = -i \beta$, with the ground (circles) and
first excited (diamonds) states each appearing as a pair. $\beta$ is increased from $10$ to $110$ in steps of $10$, and then in steps of $1$ up to $113$ for the ground states and up to $117$ for the first excited states ($r_\mathrm{min}=0.1$, $r_\mathrm{max}=100$, $N=4000$).}
\label{fig:pure_imag_pot_localized_state_energy_2}
\end{figure}

%%%%%%%%%%%%%%%%%%%%%%%%%%%%%%%%%%%%%%%%%%%%%%%%%%%%%%%%%%%%%%%%%%
\section{\label{app:proof_st}Eigenvalues of $i\hat{C}$ under $\mathcal{ST}$ symmetry}
In this appendix, we show that the eigenvalues of $i\hat{C}$ are either real or appear as complex conjugate pairs, following the argument for $\mathcal{PT}$-symmetric Hamiltonians in Ref.~\cite{PhysRevLett.80.5243}. 
Let $\ket{\psi}$ be an eigenstate of the operator $i\hat{C}\ket{\psi}=\lambda \ket{\psi}$. Applying the antilinear operator $\mathcal{ST}$ to both sides gives
\begin{equation}
    (\mathcal{ST})i\hat{C}\ket{\psi} = \lambda^{*}\left(\mathcal{ST}\ket{\psi}\right),
\end{equation}
where we use $\mathcal{T}\lambda\mathcal{T}^{-1} = \lambda^* $. Since $\left[i\hat{C},\mathcal{ST}\right]=0$, this can be rewritten as
\begin{equation}
    i\hat{C}\left(\mathcal{ST}\ket{\psi}\right) = \lambda^{*}\left(\mathcal{ST}\ket{\psi}\right).
\end{equation}
Thus, $\lambda^*$ is also an eigenvalue of $i\hat{C}$. The eigenvalues are therefore either real, when $\lambda=\lambda^*$, or appear in complex-conjugate pairs when $\lambda\neq \lambda^*$.
In the latter case, $\mathcal{ST}\ket{\psi}$ coincides with
the eigenvector of $\lambda^{*}$ up to a phase factor, if $\lambda$ is nondegenerate.

%%%%%%%%%%%%%%%%%%%%%%%%%%%%%%%%%%%%%%%%%%%%%%%%%%%%%%%%%%%%%%%%%%
\section{\label{app:beta_crit_eval}Appearance of multiple pairs of localized states and the critical values $\beta^{(\mathrm{ep})}_n$}

\begin{figure}[!t]
    \centering
    \includegraphics[width=1.0\linewidth]{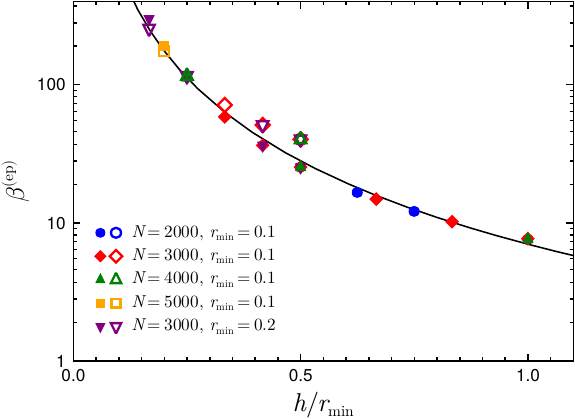}
\caption{Critical strength $\beta^{(\mathrm{ep})}_n$ of the purely imaginary potential at the exceptional point for the ground ($n=1$, filled) and first excited ($n=2$, open) states, for various values of $r_\mathrm{min}$, $r_\mathrm{max}$, and $N$, plotted as a universal function of $h/r_\mathrm{min}$. The solid curve is a fit, $\beta^{(\mathrm{ep})}_n = 7.0/(h/r_\mathrm{min})^2$ for $n=1$.}
\label{fig:pure_imag_pot_for_EP}
\end{figure}

While only one pair of localized states appears in \figref{to_pure_imag_and_localized_wavefunc}, more pairs appear for larger $|s|$, as expected from the WKB condition in \equref{n_condition_WKB}. The analytical condition in \equref{n_condition}
suggests the same tendency, although its requirement $-1<\mathrm{Re}[s]<1$ is not satisfied for large $\beta$. In \figref{pure_imag_pot_localized_state_energy_2}, we show the energies of the localized states and their trajectories as $s^2-1/4 = - i \beta$ is varied, in the same manner as in \figref{pure_imag_pot_localized_state_energy_1} but for a larger value of $\beta$ and $N$. As \equref{n_condition} now allows two localized states, two additional corresponding states with $\mathrm{Re}[2\mu E] \sim 4/h^2$ appear. As the strength of the imaginary potential is increased, they coalesce at an exceptional point at the corresponding critical strength $\beta^{(\mathrm{ep})}_n$ for each $n$.

We investigate the critical strengths $\beta^{(\mathrm{ep})}_n$ of the purely imaginary potential at the exceptional points for the ground ($n=1$) and first excited ($n=2$) states. In \figref{pure_imag_pot_for_EP}, we show the $\beta^{(\mathrm{ep})}$ for the ground (filled) and first excited (open) states for various values of $r_\mathrm{min}$, $r_\mathrm{max}$, and $N$, obtained with the same procedure as in \figref{pure_imag_pot_localized_state_energy_1}(b). Although the critical strength $\beta^{(\mathrm{ep})}$ depends on these parameters, the results for different parameter sets approximately collapse onto a single curve when plotted as a function of $h/r_{\mathrm{min}}$. 
This curve is well described by $\beta^{(\mathrm{ep})}_1 \propto (h/r_\mathrm{min})^{-2}$ for the ground state, as shown by the solid curve in \figref{pure_imag_pot_for_EP}.
The first excited state shows a slightly different behavior, so that $\beta^{(\mathrm{ep})}_1$ exceeds $\beta^{(\mathrm{ep})}_2$ only for small $h/r_\mathrm{min}$.

\bibliography{ref_TK}% Produces the bibliography via BibTeX.

\end{document}